\documentstyle[12pt]{article}
\begin{document}
\newcounter{sec}
\renewcommand{\theequation}{\thesec.\arabic{equation}}
\setcounter{sec}{2}
\def\be{\begin{equation}}
\def\ee{\end{equation}}
\def\ba{\begin{array}}
\def\ea{\end{array}}
\def\fr{\frac}
\def\la{\lambda}
\def\ep{\epsilon}
\def\c{\cal}
\def\dis{\displaystyle}
\def\pa{\partial}
\def\g{\gamma}
\def\ov{\overline}
\def\sq{\sqrt}
\parskip=6pt
\baselineskip=22pt
{\raggedleft{\sf${ ASITP}$-94-26\\}}
{\raggedleft{\sf{~~ ${ August,}$} 1994.\\}}
\bigskip\bigskip\bigskip\bigskip\bigskip
\medskip
\centerline{\Large\bf Standard Model}
\vspace{2ex}
\centerline{\Large\bf With Higgs As Gauge Field On Fourth Homotopy Group
\footnote{\sf Work
supported in part by
The National Natural Science Foundation of China.}}
\vspace{20mm}
\centerline{\large \sf Han-Ying Guo,~~Jian-Ming Li~ and~ Ke Wu }
\vspace{3.5ex}
\centerline{\sf a. CCAST (World Laboratory), P.O. Box 8730, Beijing 100080,
China;}
\vspace{1.5ex}
\centerline{\sf c. Institute of Theoretical Physics, Academia Sinica,
P.O. Box 2735, Beijing 100080, China.\footnote{\sf Mailing address.}\\}
\vspace{8ex}

\centerline{\large \sf Abstract}

\bigskip

{\it  Based upon a first principle, the generalized gauge principle, we
construct a general model with $G_L\times G'_R \times Z_2$
gauge symmetry, where $Z_2=\pi_4(G_L)$ is the fourth homotopy
group of the gauge group $G_L$,  by means of the
non-commutative differential geometry
and reformulate the Weinberg-Salam model and the standard model
with the Higgs field being a gauge field on the fourth homotopy
group of their gauge groups. We show that
in this approach not only the Higgs
field is automatically introduced on the equal footing with ordinary
Yang-Mills gauge potentials and there are no extra constraints among the
parameters at the tree level but also it
most importantly is stable against quantum correlation.}

\vfill
\newpage
\section[toc_entry]{ Introduction}

Unlike Yang-Mills gauge
fields,  Higgs fields and  Yukawa
couplings  seem to be
artificial although they play a very
important role in modern QFT.
Eventually, the price paid for them is the beauty of the gauge
principle. How to regain the beauty of the gauge principle is one of the most
intriguing problems in modern QFT.

Recently, we have generalized  the ordinary
Yang-Mills gauge theory in order to take both Lie groups and discrete
groups as gauge groups [1,2] and completed an approach to this generalized
gauge theory coupled to the fermions in the spirit of non-commutative
geometry [4, 5]. We have shown that  Higgs fields are
such  gauge fields  with respect to discrete gauge symmetry over 4-dimensional
space-time $M^4$ and the Yukawa couplings between Higgs and
fermions may automatically be introduced via generalized covariant derivatives.
In this approach, Higgs appears as discrete  fields on the equal
footing with ordinary Yang-Mills fields over spacetime $M^4$.  In other wards,
the beauty of the gauge principle may be regained. Of course,
how to understand the physical meaning of the discrete group to be gauged is a
most crucial point in this approach. On the other hand, like other approaches
[6-11] based upon the non-commutative differential geometry do not survive
the standard quantum correlation [12], the approach in [1,2] may also be
unstable against the standard quantum correlation unless there is certain
special mechanism to guarantee its stability.

In the letter [3], we have presented an $SU(2)$ generalized gauge field model
with the Higgs mechanism and shown that it is able to get rid of all those
problems based upon a first principle, the generalized gauge
principle. The key point is that we have taken into account the fourth homotopy
group of $SU(2)$ as a discrete gauge group on the footing with the Yang-Mills
gauge group $SU(2)$.
It is well known that the fourth homotopy group of $SU(2)$ is non-trivial,
$\pi_4(SU(2))=Z_2$ [13], i.e. the gauge transformations of
$SU(2)$ may be divided into two different equivalence classes.
Once the Yang-Mills fields for the gauge group
$SU(2)$ is introduced, the role played by its fourth homotopy
group must be taken into account. In view of the generalized Yang-Mills gauge
theory [1] based upon the non-commutative differential geometry, we should
also introduce the generalized gauge field with respect to this internal
discrete group $\pi_4(SU(2))$ due to the fact that the gauge transformations
depend on its elements. Although there are several remarkable advantages in
this model [3], but it is not phenomenologically realistic.

In this paper, we generalize the model presented in [3] to the realistic
cases, such as the Weinberg-Salam model and the standard model. We show that
the
most responsible internal discrete symmetry for the Higgs, say, in the standard
model is the forth homotopy group of the gauge groups,
i.e. $\pi_4(SU(3) \times SU(2)\times U(1))= \pi_4(SU(2))=Z_2$. Similar to the
model given in [3], there are several remarkable advantages in this approach.
Firstly,
it is a most natural choice of the discrete group for the Higgs and secondly
it indicates that why the Higgs in the standard model is an $SU(2)$ doublet and
$SU(3)$ singlet. Most importantly, it is stable against
quantum correlation. We will discuss these issues at the end of this paper.

In what follows, we first construct
a general model with  $G_L \times G'_R \times Z_2$ gauge symmetry,
where $Z_2$ is  the fourth homotopy
group of the gauge group $G_L\times G_R'$, i.e. $\pi_4(G_L\times G_R')$,
because we pay
our attention on the case of $\pi_4(G_R')=0$, then $\pi_4(G_L\times
G_R')=\pi_4(G_L)$.
By means of the generalized gauge theory
formulation [1] in the section 2. We also show that the Higgs
mechanism is automatically included on the equal footing with ordinary
Yang-Mills gauge fields and
there are no extra constraints at the tree level among  the
coupling constants and mass parameters under suitable normalization.
In the section 3, we reformulate the Weinberg-Salam model with Higgs being
taken as the discrete gauge field on $\pi_4((SU(2) \times U(1))=Z_2$.
Then we deal with the standard model in the section 4.
Finally, we end with some discussions and
remarks. In the Appendix, we briefly introduce the non-commutative calculus
on the discrete groups and show the Higgs fields is  the gauge
potentials with respect to the discrete gauge groups, while the Higgs potential
may be given by a Lagrangian of the Yang-Mills type.

\section[toc_entry]{ A model with  $G_L \times G'_R \times Z_2$-gauge symmetry}

Let us first construct a model of the $G_L \times G'_R \times Z_2$-gauge
symmetry, where $\pi_4(G'_R)=0$ and $Z_2$ is taken to be the fourth homotopy
group of the gauge group $G_L$, i.e. $\pi_4(G_L)=Z_2$, an intrinsic internal
discrete group of the model. Namely, the gauge transformations
of $G_L$ may be divided into two different equivalence classes. Consequently,
All leptons $\psi(x,h),
h\in Z_2$, Yang-Mills gauge potentials $A_{\mu}(x,h)$ of the gauge group
$G_L \times G'_R$ and Higgs $\Phi(x,h)$ with respect to the discrete gauge
group $\pi_4(G_L)=Z_2$ included in this model of the generalized Yang-Mills
type should be divided with respect to two elements of
$\pi_4(G_L)=Z_2$.  The construction of the model
is based upon the generalized Yang-Mills gauge theory by means of the
non-commutative differential geometry. It  combines both the Yang-Mills gauge
potentials and the
Higgs as a kind of generalized Yang-Mills gauge potentials.
For the details of this formalism, it is
referred to [1] and some relevant  notions  are briefly introduced
in the Appendix.

Let us regard those fields as elements of function space on $M^4$ as well as
on $G_L \times G'_R \times Z_2$ and assign them into two sectors
according to two elements of $\pi_4(G_L)=Z_2=\{ e, r \}$.
The $\pi_4(G_L)=Z_2$ symmetry  requires that
$$L_\mu(x,r )=UL_{\mu}(x,e)U^{-1}-\fr i g U\pa_\mu U^{-1},$$
where $U(x)$ is a topologically nontrivial gauge transformation.
Correspondingly, the left
handed fermions should also be set down at these two elements noted as
$L^e$ and  $L^r$ respectively. Namely, there is a $Z_2$ symmetry between
$L^e$ and $L^r$:
$$
 \psi(x,r)=\psi(x)^U=R_r \psi(x,e)=U(x)\psi,~~
\psi(x,e)=\psi(x)=R_r\psi(x,r)=U^{-1}\psi(x)^U$$
As for the right handed fermions, we may take $R^r=R^e=R$. Therefor, we have
\begin{equation}\begin{array}{cl}
\psi(x,e)=\psi(x)=\left(
\begin{array}{cl}
L\\
R\end{array}\right); ~~\psi(x,r)=\psi^r(x)=\left(
\begin{array}{cl}
L^r\\
R\end{array}\right)\\[6mm]
A_{\mu}(x,e)=A_{\mu}(x)=\left(
\begin{array}{ccc}
L_{\mu}&0\\[1mm]
0&R_{\mu}\end{array}\right); ~~A_{\mu}(x,r)=A^r_{\mu}(x)=\left(
\begin{array}{ccc}
L^r_{\mu}&0\\[1mm]
0&R_{\mu}\end{array}\right)\\[6mm]
\Phi(x,e)=\Phi(x)=\left(
\begin{array}{ccc}
{\frac{\mu}{\lambda}}&{-\phi}\\
{-\phi^r}^{\dag}&{\frac{\mu}{\lambda}}\end{array}\right);~~\Phi(x,r)=\Phi^r(x).
\end{array}\label {2a}\end{equation}
with the properties
$$L^r=UL, ~\phi^r=U\phi, UU^{\dag}=1,$$
$U$ is a non-trivial gauge transformation of $G_L$.
In (\ref {2a}), $L ~(R)$ is the left (right) handed fermion, $L_\mu ~ (R_\mu)$
the gauge potential valued on the Lie algebra of the gauge group $G_L ~(G'_R)$,
 $\mu$ and $\lambda$ two constants.

{}From the assignments (\ref {2a}),
it is easy to see that the field contents of the model is
of $Z_2$ symmetry and the Higgs in
such a model  may be regarded as the gauge field with respect to the gauged
$Z_2$. However, it should be mentioned that the assignments (\ref {2a})
not only assign the fields to the elements of $Z_2$ but also
imply that all fields are  arranged
into certain matrices. In fact, this aspect of the arrangements is nothing to
do with discrete gauge symmetry but  for convenience in the forthcoming
calculation. Of course, it must be kept in
mind that this is a working hypothesis and sometimes one should
 avoid certain extra
constraints coming from this working hypothesis.

{}From the general framework in [1], it follows the generalized connection
one-form
\begin{equation}
A(x,h)=A_{\mu}(x,h)dx^{\mu}+\frac{\lambda}{\mu}\Phi(x,h)
\chi,~ ~~h\in Z_2,
\end{equation}
where $\chi$ denotes ${\chi}^{r}$ in the Appendix,
and the generalized curvature two-form
\begin{equation}
\begin{array}{cl}F(h)
&=dA(h)+A(h)\otimes A(h) \\[4mm]
&=\frac{1}{2}F_{\mu\nu}(h)
{dx}^{\mu}\wedge{dx}^{\nu}+
\frac{\lambda}{\mu} F_{\mu r}(h){dx}^{\mu}\otimes{\chi}
+\frac{{\lambda}^{2}}{{\mu}^{2}}
 F_{rr}(h){\chi}\otimes{\chi}.\end{array}
\end{equation}
Using the above assignments, we get
\begin{equation}\begin{array}{cl}
F(x,e)
&=F^r(x,r)\\[4mm]
&=\frac{1}{2}\left(
\begin{array}{cc}
L_{\mu\nu}&0\\[1mm]
0&R_{\mu\nu}\end{array}\right){dx}^{\mu}\wedge{dx}^{\nu}
+\frac{\lambda}{\mu}
\left(
\begin{array}{cc}
0&{-D_{\mu}\phi}\\
-(D_{\mu}\phi^{\dag})^r&0\end{array}\right)
{dx}^{\mu}\otimes{\chi}\\
&+\frac{{\lambda}^{2}}{{\mu}^{2}}
\left(
\begin{array}{cc}
{\phi{\phi}^{\dag}-\frac{{\mu}^2}{\lambda^2}}&0\\
0&{\phi^r}^{\dag}{\phi}^r-\frac{\mu^2}{\lambda^2}\end{array}\right)
{\chi}\otimes{\chi};\\[6mm]
\end{array}\end{equation}
where
\begin{equation}
D_{\mu}{\phi}={\partial}_{\mu}{\phi}+L_{\mu}{\phi}-{\phi}R_{\mu}.
\end{equation}

Having these building blocks,
we may  introduce the generalized gauge invariant
Lagrangian with respect to each element of $\pi_4(G_L)=Z_2$, then take the Haar
integral of them over $Z_2$ to get the entire Lagrangian of the model.
Under certain consideration on the normalization in the Lagrangian, we may get
a Lagrangian without any extra constraints among the coupling constants and the
mass parameters at the tree level.

For the Lagrangian of the bosonic sector with respect to each
element of $Z_2$, we have
\begin{equation}\begin{array}{cl}
{\cal {L}}_{YM-H}(x,e)
&={\cal {L}}^r_{YM-H}(x,r)\\[4mm]
&=-\frac{1}{4N_L}Tr_L(L_{\mu\nu}L^{\mu\nu})
-\frac{1}{4N_R}Tr_R(R_{\mu\nu}R^{\mu\nu})\\[4mm]
&~+\frac{2}{N}\eta\frac{{\lambda}^{2}}{{\mu}^{2}}
Tr(D_{\mu}\phi(x))(D^{\mu}\phi(x))^{\dag}\\[4mm]
&~-\frac{2}{N}\eta^2\frac{{\lambda}^{4}}{{\mu}^{4}}
Tr(\phi(x){\phi(x)}^{\dag}-\frac{{\mu}^{2}}
{{\lambda}^{2}})^{2}+const;
\end{array}\end{equation}
where $N_L, N_R$ and $N$ are normalization constants introduced here to avoid
some extra constraints from the matrix arrangement in (\ref {2a}), $\eta$ is a
metric parameter defined by
$ \eta=< \chi, \chi >, ~Dim(\eta)={\mu}^2$.  Here we suppose
both $G_L$ and $G'_R$ be semi-simple. Eventually, this is not necessary. For
example, in the case of the Weinberg-Salam model and the standard model, $G_L$
is $SU(2)_L\times U(1)_Y$ and $SU(3)_c \times SU(2)_L\times U(1)_Y$
respectively. In those cases, we must change the way of taking normalization
in order to avoid some extra constraints from the matrix arrangement
(\ref {2a}).

For the fermionic sector, the Lagrangian with respect to each
element of $Z_2$ may also be given as follows:
\begin{equation}\begin{array}{cl}
{\cal{L}}_{F}(x,e)
&={\cal{L}}^r_{F}(x,r)\\[4mm]
&=i\overline{L}\gamma^{\mu}({\partial}_\mu+L_\mu)L
+i\overline{R}\gamma^{\mu}({\partial}_{\mu}+R_\mu)R
+\lambda (\overline{L}\phi{R}+\overline{R}{\phi}^{\dag}L).
\end{array}\end{equation}

It is easy to get the entire Lagrangian for the model:
\begin{equation}
{\cal{L}}(x)
=\frac{1}{2}\sum_{h=e, r}\{{\cal{L}}_{F}(x,h)+{\cal {L}}_{YM-H}(x,h)\}.
\end{equation}

It is easy to see that first this is a Lagrangian with the Higgs mechanism of
spontaneously symmetry breaking type included automatically which will be
studied in detail in the forthcoming sections and secondly
there do not exist any extra constraints among the coupling constants
and mass parameters which is different from other approaches [6-11].

\setcounter{sec}{3}

\section[toc_entry]{ The Weinberg-Salam Model}
\setcounter{equation}{0}

It is well known that the fourth homotopy group of the gauge group $G_L$ in the
Weinberg-Salam model is
$\pi_4(SU(2)_L\times U(1)_Y)=\pi_4(SU(2)_L)=Z_2$.
As was mentioned before, once the Yang-Mills fields for the gauge groups
$SU(2)_L\times U(1)_Y$ are introduced, the role played by their fourth homotopy
group must be taken into account. In view of the generalized Yang-Mills gauge
theory [1] based upon the non-commutative differential geometry, we should
also introduce the (generalized)
gauge field with respect to this internal discrete group $Z_2$ as well.

Now let the elements of
$\pi_4(SU(2)_L\times U(1)_Y)=Z_2$ be $\{ U_e, U_r \}$ where $ U_e$ represents
the  first topologically trivial equivalence
class of the gauge transformations  and $U_r$
the second  class which is topologically non-trivial.
We may first assign the $SU(2)_L\times U(1)_Y$
gauge fields
into two sectors with respect to these two elements as what we have done in the
last section and
make use of the formulation in the last section.
where $U(x,e)=V(x)\in U_e, U(x,r)=U(x)V(x)U^{-1}(x)\in U_e$. The bosonic part
of
the Lagrangian may also be given.

To be concrete and for the sake of simplicity, let us consider the
Weinberg-Salam model
with one family of leptons only and assign leptons, Yang-Mills gauge
potentials and Higgs into two sectors  according to
two elements of the group $\pi_4(SU_L(2)\times U_Y(1))=Z_2$ as follows:
\begin{equation}\begin{array}{cl}
\psi(x,e)=\left(
\begin{array}{cl}
L\\
R\end{array}\right);~~~
\psi(x,r)=\left(
\begin{array}{cl}
L^U\\
R\end{array}\right);\\[6mm]
A_{\mu}(x,e)=\left(
\begin{array}{ccc}
L_{\mu}&0\\[1mm]
0&R_{\mu}\end{array}\right);~~~
A_{\mu}(x,r)=\left(
\begin{array}{ccc}
L^U_{\mu}&0\\[1mm]
0&R_{\mu}\end{array}\right);\\[6mm]
\Phi(x,e)=\Phi{^{\dag}}(x,r)=\left(
\begin{array}{ccc}
{\frac{\mu}{\lambda}}&{-\phi}\\
{-\phi^U}^{\dag}
&{\frac{\mu}{\lambda}}\end{array}\right);
\end{array}\end{equation}
where $L$ and $\phi$ are $SU(2)$ doublets, $R$ an $SU(2)$ singlet and
$$\begin{array}{cl}
L(x)=\left(\begin{array}{cl}\nu_l\\
l\end{array}\right),~~~R(x)=l_R;~~~~
\phi(x)=\left(\begin{array}{cl}
{\phi}^+\\[1mm]
{\phi}^0 \end{array}\right);\\[6mm]
L_{\mu}=-ig\frac{\tau_i}{2}W^{i}_{\mu}+i\frac{g'}{2}B_{\mu},~~L^U_\mu(x,r
)=UL_{\mu}(x,e)U^{-1}-\fr i g U\pa_\mu U^{-1},~~
R_{\mu}=ig'B_{\mu}.
\end{array}$$
Thus
\begin{equation}\begin{array}{cl}
L_{\mu\nu}=
-ig\frac{\tau_i}{2}W^i_{\mu\nu}+i\frac{g'}{2}B_{\mu\nu};~~~
R_{\mu\nu}=ig'B_{\mu\nu}\\[4mm]
 D_{\mu}\phi
=({\partial}_{\mu}-ig\frac{{\tau}_{i}}{2}W^{i}_{\mu}
-i\frac{g'}{2}B_{\mu})\phi.
\end{array}\end{equation}

{}From the general model we have set up in the last section, we may directly
get
the Lagrangian. For the Yang-Mills gauge
bosons and the Higgs in the
Weinberg-Salam model, we have
\begin{equation}\begin{array}{cl}
{\cal {L}}_{YM-H}(x)
&=
-\frac{1}{4N_L}\frac{g^{2}}{2}W^{i}_{\mu\nu}{W^{i}}^{\mu\nu}-
\frac{1}{4N_Y}\frac{3g'^{2}}{2}B_{\mu\nu}B^{\mu\nu}\\[4mm]
&+\fr {2} {N} (\eta\frac{{\lambda}^{2}}{{\mu}^{2}}
Tr(D_{\mu}\phi(x))(D^{\mu}\phi(x))^{\dag}\\[4mm]
&-\fr{2}{N}\eta^2\frac{{\lambda}^{4}}{{\mu}^{4}}
(Tr\phi(x){\phi(x)}^{\dag}-\frac{{\mu}^{2}}
{{\lambda}^{2}})^{2}),
\end{array}\end{equation}
where $N_L$, $N_Y$ and $N$ are the normalization constants with respect to
$SU(2)_L$, $U(1)_Y$ and the Higgs sector.
The normalization of the coefficients of each term  results
\begin{equation}
N_L=\frac{g^{2}}{2}, ~N_Y=\frac{3g'^{2}}{2}, ~N=2\frac{{\la}^2}{{\mu}^2}\eta.
\end{equation}
It should be point out that the normalization have been taken here is
different from the one in the last section since the gauge group
$G_L=SU_L(2) \times U_Y(1)$ is not semi-simple.
Similarly, we may get the Lagrangian for leptons ${\cal L}_{F}(x)$ as follows:
\begin{equation}\begin{array}{cl}
{\cal{L}}_{F}(x)
&=-i\overline{L}(x){\gamma}^{\mu}(
{\partial}_{\mu}+ig\frac{\tau_i}{2}W^i_\mu-
i\frac{g'}{2}B_{\mu}){L}\\[4mm]
&~~~-i\overline{R}(x){\gamma}^{\mu}
({\partial}_{\mu}-ig'B_{\mu}){R}\\[4mm]
&~~~-\lambda (\overline{L}(x)\phi(x)R(x)
+\overline{R}(x){\phi(x)}^{\dag}L(x)).
\end{array}\end{equation}
Thus, the entire Lagrangian for the  Weinberg-Salam model reads
\begin{equation}
{\cal{L}}(x)={\cal L}_{F}(x)+{\cal L}_{YM-H}(x).
\end{equation}

It is easy to see that the Higgs potential takes its minimum value at
 $Tr(\phi{\phi}^{\dag})=(\frac{\mu}{\lambda})^{2}$ and
the continuous gauge symmetry will spontaneously be broken down when
 the vacuum expectation value (VEV) is taken as
\begin{equation}
<{\phi}>=\left(
\begin{array}{cl}
0\\[1mm]
\frac{\rho_0}{\sqrt{2}}
\end{array}\right),
\end{equation}
where $\rho_0=\sqrt{2}\frac{\mu}{\lambda}$.
Now we take the VEV of $\phi$ and introduce a new field
$\eta(x)$ as the  Higgs field in the model
\begin{equation}
{\phi}=\left(
\begin{array}{cl}
0\\[1mm]
\frac{\rho_0+\rho(x)}{\sqrt{2}}\end{array}\right),
\end{equation}
as well as the photon and $Z$ boson via $W$ bosons and the Weinberg angle
\begin{equation}\begin{array}{c}
A_{\mu}=B_{\mu}cos{\theta}_W+W_{\mu}^3sin{\theta}_W\\
Z_{\mu}=B_{\mu}sin{\theta}_W-W_{\mu}^3cos{\theta}_W\\
g\sin{\theta}_{W}=g'\cos{\theta}_{W}=\frac{gg'}{\sqrt{g^{2}
+g'^{2}}}=e,
\end{array}\end{equation}
where $e$ is the charge of the positron.
Using these definitions, we get
\begin{equation}
sin^2{\theta}_{W}=\frac{g'^{2}}{g^{2}+g'^{2}}~~\{=\frac{N_Y}{3N_L+N_Y}\}.
\end{equation}
And we have
\begin{equation}\begin{array}{cl}
&Tr\{D_{\mu}\phi(D_{\mu}\phi)^{\dag}-
\eta\frac{\la^2}{\mu^2}(\phi{\phi}^{\dag}
-\frac{{\mu}^{2}}{{\lambda}^{2}})^{2}\}\\[4mm]
&=\frac{1}{2}{\partial}_{\mu}\rho{\partial}^{\mu}\rho+\frac{g^{2}}{4}
(\rho_0+\rho)^{2}W_{\mu}^{-}W_{\mu}^{+}+\frac{1}{8}(g^{2}+g'^{2})
(\rho_0+\rho)^{2}
Z_{\mu}Z_{\mu}\\[4mm]
&~~~-\eta\frac{\la^2}{\mu^2}{\rho}^{2}(\rho_0^{2}+\rho_0\rho+\frac{{\rho}^{2}}{4})+const.
\end{array}\end{equation}
It is easy to see that only
$A_{\mu}$ and $\nu_l$ remain massless while fermion $l$ together with
$W^{\pm}$ and Z
become massive and  following mass relations hold at the tree level:
\begin{equation}\begin{array}{cl}
&M_{fermion}=\mu,~~~M_{W}=\frac{1}{2}g\rho_0,\\[4mm]
&M_{Z}=\frac{1}{2}\sqrt{g^{2}+g'^{2}}\rho_0=M_{W}/\cos{\theta}_{w},\\[4mm]
&M_{Higgs}=2\sq\eta.
\end{array}\end{equation}

It is easy to see that all these relations at the tree level are the
same as the ones for the Weinberg-Salam model except the last one for the Higgs
mass but different from what is  given in
[6]. The reason is that we have introduced two independent normalization
constants $N_L$, $N_Y$  and $N$ in order to avoid some extra constraints from
the matrix arrangement (1). In fact, if we would take $N_L=N_Y$ we could get
the same constraint for the Weinberg angle in [6]. In other wards, as was
mentioned in [1]
the constraints in [6] are not
essential but
completely dependent on the working hypothesis. As for the Higgs mass given
here
at the tree level, it depends on the metric parameter $\eta$. If we let it free
of choice, there is no constraint for the Higgs mass at all.

\setcounter{sec}{4}
\section[toc_entry]{ Standard Model for Electroweak-Strong Interaction\\}
\setcounter{equation}{0}
We now turn to the  standard model for the electroweak-strong
interactions. This should be more realistic from both conceptual and
phenomenological points of view. We take into account the colour
degree of freedom together with the weak isospin and the weak hypercharge
degrees of
freedom for both leptons and quarks in three families. Similar to what we
have done
in the last section, we first introduce the gauge fields of the Yang-Mills
gauge groups $SU(2)_L\times U(1)_Y \times SU(3)_c$ and assign them
into two elements of $\pi_4(SU(2)_L\times U(1)_Y \times SU(3)_c)=Z_2$ symmetry
respectively. Such assignments for the fermions and the Higgs will be given
according to their couplings to the Yang-Mills gauge fields as well.

The assignment for the fermions  with respect to
$\pi_4(SU(2)_L\times U(1)_Y \times SU(3)_c)=Z_2=\{ e, r \}$
may be taken as follows:
\be\ba{cl}
\psi(x,e)=\left(\ba{cl}
L\\R\ea\right),~~~
\psi(x,r)=\left(\ba{cl}
L^{\cal U}\\R\ea\right),\ea\ee
with $L^{\cal U}={\cal U}L=\left(\ba{cc}U \otimes I_3\otimes I_3^c&\\&U\times
I_3\ea\right)L$, U the topologically
non-trivial gauge transformation of $SU(2)_L$ and
\be
L=\left(
\begin{array}{cl}
u^{c}\\
\vdots\\
b^{c}\\
{\nu}_{e}\\
\vdots\\
\tau\end{array}\right)^{GFC}_{L}
,~~~~R=\left(
\begin{array}{cl}
u^{c}\\
\vdots\\
b^{c}\\
e\\
\mu\\
\tau\end{array}\right)_{R}. \label{c}\ee
Here superscript $c$ stands for the colour degree of freedom.
 Taking into account all strong and
electroweak interactions among leptons and quarks according to (\ref c),
we assign the gauge fields as follows:
$$A_{\mu}(x,e)=\left(
\begin{array}{ccc}
L_{\mu}&0\\[1mm]
0&R_{\mu}\end{array}\right),~~
A_{\mu}(x,r)=\left(
\begin{array}{ccc}
L^{\cal U}_{\mu}&0\\[1mm]
0&R_{\mu}\end{array}\right),$$
where
\be
\ba{cl}
L_\mu&=-\left(\ba{ccc}\frac{ig}{2}{\tau}^{I}_{i}W_{\mu}^{i}\otimes I_3^G
\otimes I_3^C&\\[4mm]
&&\frac{ig}{2}{\tau}^{I}_{i}W_{\mu}^{i}\otimes I_3^G
\ea\right)\\[6mm]
&-igB_\mu\left(\ba{ccc}
\frac 1 6 I_2\otimes I_3^G\otimes I_3^C&\\
&&-\frac 1 2 I_2\otimes I_3^G\ea\right)
-\left(\begin{array}{ccc}
I_{2}\otimes I_3^G\otimes
\frac{ig_{2}}{2}G^{i}_{\mu}
{\lambda}^{C}_{i}&\\[3mm]
&&0
\end{array}\right),\\[10mm]
L^{\cal U}_\mu&={\cal U}L_\mu {\cal U}^{-1}+{\cal U}\pa_\mu {\cal
U}^{-1}\\[6mm]
R_\mu&=-ig_{1}B_{\mu} \left(\ba{ccc}\left(\begin{array}{ccc}
\frac{2}{3}\\
&&{-\frac{1}{3}}\ea\right)\otimes I_3^G\otimes I_3^C\\[4mm]
&&-I_3^G\ea\right)
-\left(\begin{array}{ccc}
I_{2}\otimes I_3^G\otimes
\frac{ig_{2}}{2}G^{i}_{\mu}{\lambda}^{C}_{i}&\\[3mm]
&&0
\end{array}\right),
\end{array}\ee
where $G^{i}_{\mu}, i=1, \cdots, 8,$ are gluons, ${\lambda}_{i}$
$3\times 3 $
Gell-Mann matrices, and $I_{n}$ $n\times n$ unit matrices.
For the  Higgs
field, we take it as before
\be\Phi(x,e)=\Phi^{\dag}(x,r)=\left(
\begin{array}{ccc}
0&{\Phi(x)}\\
({\c U}\Phi(x))^{\dag}&0\end{array}\right).\ee
But, $\Phi(x)$ field
being gauge field with respect to $Z_{2}$-symmetry is more complicated:
$$\Phi(x)=\left(\ba{ccc}
\left(\begin{array}{ccc}
{\phi}^{0*}&{\phi}^{+}\\[1mm]
-{\phi}^{+*}&{\phi}^{0}\end{array}\right)\otimes I^{G}_{3}\otimes
I_3^C&\\[6mm]
&\left(\ba{c}
\phi^{\dag}\\
\phi^0\ea\right)\otimes I_3^G\ea\right)
\cdot\left(\begin{array}{ccccc}
\lambda^{U}\otimes I_3^C\\[1mm]
&\lambda^{D}\otimes I_3^C\\[1mm]
&&\lambda^{L}\end{array}\right).$$
Here the blocks in the last matrix are $3\times 3$  matrices in the
space of generation, $\lambda^U, \lambda^D$ the matrices for quarks
and $\lambda^L$ the matrices for leptons. These matrices play the role of the
Yukawa coupling constants.

Now we may write down the generalized connection one-form including both
ordinary Yang-Mills potentials and the Higgs field on the equal footing
 and the generalized curvature two-form.
Especially, the components $F_{\mu r}$ of the generalized field strength are
the ordinary covariant derivative of the Higgs field as before:
$$D_{\mu}\Phi={\partial}_{\mu}\Phi+
L_\mu\Phi-\Phi R_\mu$$

Making use of the model in the section 2, we may get the Lagrangian as long as
the normalization is suitably taken. The bosonic part of the entire gauge
invariant Lagrangian, by some
straightforward calculation, is
\be\begin{array}{cl}
{\cal L}_{YM-H}&=-\frac 1 N <F,\overline{F}>\\[4mm]
&=-\frac{1}{4N_L}6g^{2}W^{i\mu\nu}W^{i\mu\nu}
-\frac{1}{4N_Y}10g_{1}^{2}B_{\mu\nu}B^{\mu\nu}
-\frac{1}{4N_c}6g_{2}^{2}G^{i}_{\mu\nu}G^{i\mu\nu}\\[4mm]
&~~+\frac 1 N
\{2{\eta}\frac{{\sigma}_{1}'}{{\mu}^{2}}(D_{\mu}\pi)^{\dag}D^{\mu}\pi-
2{\eta}^2\frac{{\sigma}_{2}'}{{\mu}^{4}}({\pi}^{\dag}\pi-
\frac{{\sigma}_{1}}{{\sigma}_{2}}{\mu}^{2})^{2}\},
\end{array}\ee
where  $N_L$, $N_Y$, $N_c$ and $N$ are normalization constants with respect
to gauge fields W, B, G and the Higgs sector respectively, the field $\pi$ is
introduced as $\pi=\left(\ba{c}
\phi^{\dag}\\
\phi^0\ea\right)$, and
\be\begin{array}{cl}
&{\sigma}_{1}=Tr\left(\begin{array}{ccccccc}
\lambda^U{\lambda^U}^{\dag}\otimes I^{C}_{3}&\\[1mm]
&&\lambda^D{\lambda^D}^{\dag}\otimes I^{C}_{3}&\\[1mm]
&&&&\lambda^L{\lambda^L}^{\dag}\end{array}\right),\\[10mm]
&{\sigma}_{2}=Tr\left(\begin{array}{ccccccc}
\lambda^U{\lambda^U}^{\dag}\otimes I^{C}_{3}&\\[1mm]
&&\lambda^D{\lambda^D}^{\dag}\otimes I^{C}_{3}&\\[1mm]
&&&&\lambda^L{\lambda^L}^{\dag}
\end{array}\right)^{2}.\end{array}
\ee
The normalization of the coefficients of the terms in the entire Lagrangian
leads to that
\be
N_L=6g^{2},~~N_Y=10g_{1}^{2},~~N_c
=6g_{2}^{2},~~N=2\frac{{\sigma}_{1}}{{\mu}^{2}}
{\eta}.\ee
This gives rise to the following form for the Yang-Mills-Higgs Lagrangian
\be\begin{array}{cl}
{\cal L}_{YM-H}=&-\frac{1}{4}W^{i}_{\mu\nu}W^{i\mu\nu}
-\frac{1}{4}B_{\mu\nu}B^{\mu\nu}
-\frac{1}{4}G^{i}_{\mu\nu}G^{i\mu\nu}\\[4mm]
&+(D_{\mu}\pi)^{\dag}D^{\mu}\pi-
\frac{\eta}{\mu^2} \frac{{\sigma}_{2}}{{\sigma}_{1}}({\pi}^{\dag}\pi-
\frac{{\sigma}_{1}}{{\sigma}_{2}}{\mu}^{2})^{2}.
\end{array}\label {l}\ee
It is obviously  that together with the Lagrangian of the usual gauge fields
 the kinetic energy of Higgs field and the interaction between Higgs field
and the usual gauge fields are all included here.

It is easy to see that when $\pi$ field takes value
\be
|\pi|=\sqrt{\frac{{\sigma}_{1}}{{\sigma}_{2}}}{\mu},\ee
the Higgs potential is at its minimum. If we set the vacuum expectation value
 as
\be
<\pi>=\left(\begin{array}{cl}
0\\[1mm]
\frac{\rho_0}{\sqrt{2}}\end{array}\right),~~v=
\sqrt{\frac{2{\sigma}_{1}}{{\sigma}_{2}}}
\mu,\ee
the continuous gauge symmetry is broken down.
Introducing a field $\rho(x)$
\be
\pi=\left(\begin{array}{cl}
0\\
\frac{\rho_0+\rho(x)}{\sqrt{2}}\end{array}\right),\ee
and the photon A as well as the boson Z
\be\ba{cl}
&A_{\mu}=\sin{{\theta}_{w}}W^{3}_{\mu}+\cos{{\theta}_{w}}B_{\mu},\\[4mm]
&Z_{\mu}=\cos{{\theta}_{w}}W_{\mu}^{3}-\sin{{\theta}_{w}}B_{\mu},\ea\ee
where
\be
g\sin{{\theta}_{w}}=g'\cos{{\theta}_{w}}=\frac{gg'}{\sqrt{g^{2}
+g'^{2}}}=e,\ee
we may get spontaneous symmetry breaking version of (\ref l).

For the fermions, we can also write down the Lagrangian in a way similar to
what we have done for the model of leptons in the
last subsection:
\be
\begin{array}{cl}
{\cal L}_{F}(x)&={\displaystyle\sum}_{i}\overline{q_{i}}
i{\gamma}^{\mu}{D_{\mu}}q_{i}
+{\displaystyle\sum}_{i}\overline{l_{i}}
i{\gamma}^{\mu}{D}_{\mu}l_{i}\\[4mm]
&~~-\left\{\left(\bar e_R~\bar \mu_R~\bar \tau_R\right)\lambda^L
\left(\ba{c}e_L\\
\mu_L\\
\tau_L\ea\right)
+\left(\bar u_R^c~\bar c_R^c~\bar t_R^c\right)\lambda^U\otimes I_3
\left(\ba{c}u_L^c\\
c_L^c\\
t_L^c\ea\right)\right.\\[4mm]
&~~\left.+\left(\bar d_R^c~\bar s_R^c~\bar b_R^c\right)\lambda^D\otimes I_3
\left(\ba{c}d_L^c\\
s_L^c\\
b_L^c\ea\right)+h.c. \right\}\frac{\rho_0}{\sqrt{2}}(1+\frac \rho {\rho_0})
\ea\ee
As is well known, both $\lambda^L$ and $\lambda^U$ may be  diagnolized as
$$\lambda^L=\left(\ba{ccc}\lambda_e&&\\&\lambda_\mu\\&&\lambda_\tau\ea\right),~~
\lambda^U=\left(\ba{ccc}\lambda_u&&\\&\lambda_c\\&&\lambda_t\ea\right)$$
while $\lambda^D$ may be written as
\be
\lambda^D=V\left(\ba{ccc}\lambda_d&&\\&\lambda_s\\&&\lambda_b\ea\right)V^{\dag},\ee
where $V$ is the Kobayashi-Maskawa matrix.

Since the  mass of top quark is much
heavier than other fermions, i.e. $m_{t}\gg m_{i}$, where $m_i$ is the
mass for the fermion $i$ except $t$,
we have
\be
\frac{{\sigma}_{2}}{{\sigma}^{2}_{1}}=\frac{1}{3},
{}~~~~~\frac{{\sigma}_{1}}{{\sigma}_{2}}
=\frac{1}{\lambda^2_{t}},~~~~~\epsilon=\sqrt{ \frac{\eta}{\mu^2}}
\ee
where  $\lambda_{t}$ is the coupling constant corresponding to the  top quark.
Then the Lagrangian for the generalized gauge fields can be rewritten as
\be\begin{array}{cl}
{\cal L}_{YM-H}=&-\frac{1}{4}W^{i}_{\mu\nu}W^{i\mu\nu}
-\frac{1}{4}B_{\mu\nu}B^{\mu\nu}
-\frac{1}{4}G^{i}_{\mu\nu}G^{i\mu\nu}\\[4mm]
&+(D_{\mu}\pi)^{\dag}D^{\mu}\pi-
\eta\frac {\lambda^2_t}{\mu^2}  ({\pi}^{\dag}\pi-
\frac{{\mu}^{2}}{\lambda_{t}^2})^{2}.
\end{array}\label{e}\ee
Consequently, when $\pi$ field takes value $|\pi|=\frac{\mu}{\lambda_{t}}$,
the Higgs potential
is at its minimum. If we set
$$<\pi>=\left(\begin{array}{cl}
0\\
\frac{\rho_0}{\sqrt{2}}\end{array}\right),~~~\rho_0=\frac{\sqrt{2}}
{\lambda_{t}}\mu$$
the symmetry $SU(2)_L \times U(1)_Y$ will spontaneously be broken
down.
Introducing new field $\rho$ to replace the field $\pi$ in eq.(\ref e),
just as we have done in the last section, and adding the fermionic part
through generalized covariant
derivative, we get the final expression of the entire  Lagrangian as follows
\be
\begin{array}{cl}
{\cal L}(x)&={\cal L}_{F}(x)+{\cal L}_{YM-H}(x)\\[4mm]
&={\displaystyle\sum}_{i}\overline{q_{i}}
i{\gamma}^{\mu}{D_{\mu}}q_{i}
+{\displaystyle\sum}_{i}\overline{l_{i}}
i{\gamma}^{\mu}{D}_{\mu}l_{i}\\[4mm]
&-\left\{\left(\bar e_R~\bar \mu_R~\bar
\tau_R\right)\left(\ba{l}m_e\\~~~~m_\mu\\~~~~~~~~m_\tau\ea\right)
\left(\ba{c}e_L\\
\mu_L\\
\tau_L\ea\right)
+\left(\bar u^c_R~\bar c^c_R~\bar
t^c_R\right)\left(\ba{l}m_u\\~~~~m_c\\~~~~~~~~m_t\ea\right)\otimes I^c_3

\left(\ba{c}u^c_L\\
c^c_L\\
t^c_L\ea\right)\right.\\[4mm]
&\left.+
\left(\bar d^c_R~\bar s^c_R~\bar
b^c_R\right)V\left(\ba{l}m_d\\~~~~m_s\\~~~~~~~~m_b\ea\right)V^{\dag}\otimes
I^c_3\left(\ba{c}d^c_L\\
s^c_L\\
b^c_ L\ea\right)+h.c.\right\}(1+\frac \rho {\rho_0})
\\[4mm]
&-\frac{1}{2}W^{\dag}_{\mu\nu}W^{\mu\nu}-\frac{1}{4}Z_{\mu\nu}
Z^{\mu\nu}
-\frac{1}{4}A_{\mu\nu}A^{\mu\nu}\\[4mm]
&+\frac{1}{2}{\partial}_{\mu}{\rho}{\partial}^{\mu}{\rho}+\frac{g^{2}}{4}
(\rho_0+\rho)^{2}W^{-}W^{+}+\frac{g^2}{8{\cos{{\theta}_{w}}}^{2}}
(\rho_0+\rho)^{2}Z_{\mu}Z^{\mu}\\[4mm]
&-\frac{\eta}{\mu^2}
\frac{{\sigma}_{2}}{{\sigma}_{1}}(\rho_0^{2}{\rho}^{2}+\rho_0{\rho}^{3}
+\frac{{\rho}^{4}}{4})\end{array}.\ee

It is easy to see that  neutrinos, photon and gluons remain massless while
other particles become massive. And we can also get the following
mass relations,
\be
\begin{array}{cl}
&m_{W}=\frac{1}{2}g\rho_0\\[4mm]
&m_{Z}=\frac{m_{W}}{\cos{{\theta}_{w}}}\\[4mm]
&m_{Higgs}\approx 2\sqrt{\eta}\\[4mm]
&m_{t}\approx  \mu
\end{array}.\ee

Similar to the last section, it is easy to see that all these relations at the
tree level are the
same as the ones for the standard  model except that for the Higgs
mass. The Higgs mass given here
also depends on the metric parameter $\eta$. If we let it free
of choice, there is no constraint for it at all. Otherwise, if we would take
\be
\eta=\mu^2\ee
we could get
\be
M_H=2m_t.\ee
However, there is no profound reason to do so.

\section[toc_entry]{ Concluding Remarks}

Now we summarize what we have done  as follows:

We have first constructed a general model with $G_L\times G'_R \times Z_2$
gauge symmetry, where $Z_2$ is $\pi_4(G_L)$,
by means of the generalized gauge theory on both Lie groups and
discrete groups. We have shown that the Higgs
mechanism is automatically included in the generalized gauge theory and there
are no constraints among the parameters at the tree level in this model.
Then we have reformulated the Weinberg-Salam model and the standard model
with the Higgs field being a gauge field with respect to  the fourth homotopy
group of the gauge groups, i.e.
$\pi_4(SU(3) \times SU(2)\times U(1))= \pi_4(SU(2))=Z_2$.

It is worthy to point out that there are several advantages in this approach
as was mentioned at the introduction of this paper.
First of all, this $\pi_4(SU(2))=Z_2$ symmetry is a most natural internal
symmetry to be gauged  in these models
in the sense of non-commutative differential calculus on the function
space on $M^4$ as well as on $G_L \times G'_R \times \pi_4(G_L\times
G'_R)=Z_2$.
In fact, for these models, the fourth homotopy group of the gauge groups is
already there
and it should play certain role in the gauge theory. What we have
done here is just to combine the ordinary Yang-Mills gauge theory with the
non-commutative differential calculus in the function
space on this discrete group to formulate a generalized gauge theory with Higgs
and spontaneously symmetry breaking. In other wards, the Higgs mechanism should
be introduced automatically at same footing with the ordinary Yang-Mills gauge
field theory, if the role played by the fourth homotopy group of the gauge
groups would be taken into account  at very beginning.

Secondly, it is also interesting  to see that the  mystery of the
Higgs pattern in the standard model may be understood better. In fact, that
$\pi_4(SU(3))=0$, $\pi_4(U(1))=0$ and $\pi_4(SU(2))=Z_2$ indicates that
Higgs should play certain role for the $SU(2)$ gauge field and nothing to do
for the $SU(3)$ gauge symmetry. Taking into account the properties of the
fermions the Higgs in the standard model should be an $SU(2)$ doublet and
$SU(3)$ singlet.

Finally, it is remarkable that the
approach presented here with the fourth homotopy group of the gauge groups
being
the discrete gauge group is stable against quantum correlation. This is due to
the following reasons. Firstly, there are no constraints among the parameters
at
the tree level so that we do not need to pay attention to them in the
course of quantization and renormalization. Secondly,
since the Higgs potential is automatically introduced in the
generalized gauge theory, the $SU(2)$ gauge symmetry should be spontaneously
broken down. Therefore, this $Z_2$ symmetry, the fourth homotopy group of the
gauge symmetry in those models is also broken down
as well. Consequently, what we got is, say, the same version as the
ordinary standard model and we of course do not need to concern about this
$Z_2$-gauge symmetry when we consider the quantum correlation of the model.
Needless to say, this is a very
important point different from other approaches to the Higgs by means of the
non-commutative differential geometry. In fact, Connes like approaches [6-10]
do not survive the quantum correlation [11].

In conclusion, the Higgs mechanism may be a part of a generalized Yang-Mills
gauge theory as long as a global aspect, the fourth homotopy group, of the
gauge group is taken into account in the sense of the non-commutative
differential geometry. For the standard model the most natural and meaningful
discrete symmetry on which the Higgs is  a  generalized gauge
fields  is just the fourth homotopy group of the gauge groups.

It is clear that the model presented in the section 2 may be generalized to
the case of $ \pi_4(G_L \times G'_R)=Z_2 \times Z_2$ and it may be applied to
the left-right symmetric model. On the other hand, since the fourth homotopy
group of $SU(5)/ ( SU(3) \times SU(2) \times U(1))$ is also non-trivial, it may
play certain role in the SU(5)-GUT together with the
fourth homotopy group of  $ SU(3) \times SU(2) \times U(1)$. And all models of
these kind may have the same advantages as the approach presented in this
paper.
Especially, all of them may be stable against quantum correlation.
As for other discrete symmetries such as $CPT$ and so on, they may play other
roles such as CP violation and so on. We will study these issues elsewhere.

\bigskip
\bigskip

\setcounter{sec}{0}
\section[toc_entry]{ Appendix}
\setcounter{equation}{0}
\centerline{\large\bf  Differential Calculus on Discrete Group G}

In this appendix, we briefly introduce some notions in the non-commutative
differential
calculus on the function space on discrete groups and show the Higgs may be
regarded as the (generalized) gauge potential on the gauged discrete group.
For the details, it is
referred to Sitarz in [5] and our papers [1].

Let G be discrete group of size $N_G$, its elements are $\{e,g_1, g_2,\cdots,
g_{N_G-1}\}$,
and ${\c A}$ the algebra of the all complex valued functions on G.
In order to construct the first
order differential calculus $(\Omega^1, d)$, one can give first the definition
of its dual space
${\c F}$, the vector space on ${\c A}$ with basis $\partial_i$, $(i=1,\cdots,
N_G-1)$ as follows:
\be
{\partial}_{g}=f-R_{g}f,~~g\in G',f\in {\cal A}, \ee
where
\be (R_{i}f)(g)=f(g\odot g_i)\ee
which is nothing but the difference operator on ${\c A}$, and satisfies

\be
{\partial}_{i} {\partial}_{j}={{\displaystyle\sum}_{k}}C_{ij}^{k}
{\partial}_{k}, ~~C_{ij}^k=\delta^k_i+\delta^k_j-\delta^k_{i\cdot j}\ee
where $i,j,\cdots, (i\cdot j)$ denote $g_i,g_j,\cdots, (g_i\cdot g_j)$
respectively.
The basis $\chi^i$ of $\Omega^1$ are just the dual of $\partial_i$,
\be
 {\chi}^{i}({\partial}_{j})={{\delta}^{i}}_{j}. \ee
Then the first order differential calculus $(\Omega^1,d)$ is given by
\be
df=\dis{\sum^{N_G-1}_{i=1}}\partial_i f \chi^i \ee

For the differential algebra $\Omega^*$ over ${\c A}$ and exterior derivative,

\be
d:~{\Omega}^{n} \rightarrow {\Omega}^{n+1}\ee
satisfies the nilpotency and the graded Leibniz rule
\be\begin{array}{cl}
(i) ~~&d^{2}=0,\\[4mm]
(ii)~~&d(fg)=df\cdot g+(-1)^{degf}f\cdot dg,~~~~~\forall f, g \in {\Omega}^{*},
\end{array}\ee
could be obtained provided that $\chi^i$ satisfy the following  two conditions,
\be
\begin{array}{cl}
{\chi}^{i}f=&(R_{i}f){\chi}^{i},~~f\in {\cal A},\\[4mm]
d{\chi}^{i}=&-{\sum}_{j,k}C^{i}_{jk}{\chi}^{j}\otimes {\chi}^{k},
{}~~g\in G'.\end{array}\ee

 The involution operator $*$ on the differential algebra  $\Omega^*$ is well
defined if it
 agrees with the complex conjugation on ${\cal A}$,
takes the assumption that $({\chi}^{g})^{*}=-{\chi}^{g^{-1}}$,
and (graded)
commutes with d, i.e. $d({\omega}^{*})=(-1)^{deg\omega}({d\omega})^{*}$.
The integral, which remains  invariant under the group
action,  is introduced as a complex valued linear functional
on ${\cal{A}}$ as,
\be
{\int}_G f=\frac{1}{N_{G}}\dis{\sum_{g\in G}}f(g).\ee

Let us consider the case that there are Lie group transformations among the
elements of the function
space and those transformations also depend on the elements of the discrete
group. Then the derivatives introduced above are no longer covariant. In order
to
get meaningful differential calculus in this case, the connection one form is
needed to define the covariant exterior differential:
\be
D=d_G+\phi,
\ee
where the  connection one form $\phi$ may be written as
\be
\phi=\phi_g\chi^g\ee
from which we get the generalized curvature two form
\be
F=A+A\otimes A=\dis{\sum_{g,h}}F_{gh}\chi^g\otimes\chi^h \ee
where
\be
F_{gh}=\pa_g\phi_h+\phi_gR_g\phi_h-C{_gh}^k\phi_k.\ee
 This formula is simpler in terms of $\Phi=1-\phi$
\be
F_gh=\\Pgi_gR_g(\Phi_h)-\Phi_{h\otimes g}.\ee
After introducing the the metric, we can get the Lagrangian for the theory.

For the $Z_2$ case, we can define the metric as
\be
<\chi, \chi>=\eta,~~<\chi\otimes\chi, \chi\otimes \chi>=\eta^2\ee
then
\be
{\c L}=-<\ov F,F>=-\eta^2(\Phi\Phi^{\dag}-1)^2.\ee
This is of Higgs potential type up to some coupling constants.
To get the entire Lagrangian of the Higgs, we need to consider the space-time
part. For detail it is  referred to [1, 5].

\centerline{\large \sf {References}}

\bigskip

\begin{enumerate}

\item H.G. Ding, H.Y. Guo, J.M. Li and K. Wu,
Comm. Theor. Phys. {\bf 21} (1994) 85-94;
Higgs as gauge  fields on discrete
groups and standard models for electroweak and electroweak-strong interactions;
To appear in Z.Phsik. {\bf C}.

\item
H.G. Ding, H.Y. Guo, J.M. Li and K. Wu,
 J. Phys.  {\bf A 27} (1994) L75-L79;  ibid. L231-L236.

\item H.Y. Guo and J.M. Li, An $SU(2)$ generakized gauge field model with Higgs
mechanism; Preprint ASITP-94-40; July, 1994.

\item A. Connes, \-{\it Non-Commutative \- Geometry}~~
\-English \- translation  of \- Geometrie \- Non-Commutative, IHES \- Paris,
 Interedition.

 \item
 A. Sitarz,  Non-commutative Geometry and  Gauge Theo\-ry on Dis\-crete
Groups, preprint {\bf TPJU}-7/1992.

\item A. Connes, in: The Interface of  Mathematics and Particle Physics,\\
eds. D. Quillen, \- G. Segal and S. Tsou \- (Oxford U. P, Oxford 1990).

\item A. Connes and J. Lott, \- Nucl. Phys. (Proc. Suppl.) {\bf B18}, 44
(1990).

\item A. Connes and Lott, \- Proceedings of 1991 \-Cargese Summer \-
Conference.

\item D. Kastler, Marseille, CPT preprint {\bf CPT-91}/P.2610, {\bf
CPT-91}/P.2611.

\item A. H. Chamseddine, \- G Felder and J. Fr\"ohlich, \- Phys. Lett. \-
 {\bf 296B} (1993) 109, \- Zurich preprint  \- {\bf ZU-TH}-30/92 and
Zurich \- preprint \- {\bf ETH-TH}/92/44.
 A. H. Chamseddine \- and J. Fr\"ohlich, \- $SO(10)$ \- Unification in \-
 Noncommutative
\- Geometry, {\bf ZU-TH}-10/1993.

\item R. Coquereaux, G. Esposito-Far\'ese and G Vaillant, Nucl Phys
{\bf B353}  689 (1991).

\item E. \'Alvarez, J.M. Gracia-Bondia and C.P. Martin,
Phys. Lett. {\bf B306}, 53 (1993).

\item S.T. Hu, Homotopy Theory, Academic Press, New York, 1959.

\end{enumerate}

\end{document}